\documentclass{optica-article}

\journal{opticajournal} 

\articletype{Research Article}

\usepackage{lineno}
\usepackage{array} 
\usepackage{changepage}
\usepackage{setspace}
\usepackage{wrapfig}
\usepackage{tabularx}
\usepackage{amsmath}

\begin{document}

\title{Nonlinear Symmetry Breaking to Enhance the Sagnac Effect in a Microresonator Gyroscope}

\author{Thariq Shanavas,\authormark{1,$^\dagger$} Gregory Krueper,\authormark{1,$^\dagger$}, Jiangang Zhu,\authormark{2} Wounjhang Park,\authormark{3,4} and Juliet T. Gopinath\authormark{1,3,4,*}}

\address{
\authormark{$^\dagger$}These authors contributed equally to this work\\
\authormark{1}Department of Physics, University of Colorado, Boulder, CO, USA\\
\authormark{2}DeepSight Technology, Santa Clara, CA, USA \\
\authormark{3}Department of Electrical, Computer and Energy Engineering, University of Colorado, Boulder, CO, USA\\
\authormark{4}Materials Science Engineering Program, University of Colorado, Boulder, CO, USA\\
}

\email{\authormark{*}julietg@colorado.edu} 


\begin{abstract*}
Optical gyroscopes based on the Sagnac effect have been widely used for inertial navigation in aircrafts, submarines, satellites and unmanned robotics. With the rapid progress in the field of ultrahigh-quality whispering gallery mode and ring resonators in recent years, these devices offer the promise of a compact alternative to ring-laser gyroscopes (RLGs) and fiber-optic gyroscopes (FOGs). Yet, successful commercialization of a microresonator gyroscope has been hindered by the scaling of the Sagnac effect with resonator area. While several techniques have been proposed to enhance the Sagnac effect in microresonators, these enhancements also amplify the thermal noise in the microresonator. Here, we present a novel approach to measuring the Sagnac signal in chip-scale devices that overcomes this fundamental noise limitation to achieve unprecedented performance in a 200~\textmu m optical resonator - the smallest reported to date. Our proof-of-concept design shows a $10^4$ enhancement of the Sagnac signal while simultaneously suppressing thermal noise by 27 dB and environmental contributions to noise by 22 dB. We believe this approach offers a pathway for integrated photonic gyroscopes with sensitivities that match or exceed RLGs and FOGs.
\end{abstract*}

\section{Introduction}




Optical gyroscopes, leveraging the Sagnac effect \cite{post1967sagnac}, have found diverse applications across numerous fields demanding high precision rotational sensing. Their key characteristics of no moving parts, vibration tolerance, high sensitivity and immunity to electromagnetic interference offer significant advantages over mechanical gyroscopes in demanding operational environments \cite{yang2025micro, arianfard2023sagnac}. Consequently, Sagnac interferometers including ring laser gyroscopes (RLGs) \cite{chow1985ring}, fiber optic gyroscopes (FOGs) \cite{bergh1984overview}, and hemispherical resonator gyroscopes (HRGs) \cite{rozelle2009hemispherical} dominate the market for navigation-grade instruments in aircrafts, submarines, and autonomous robotic systems today \cite{oho2002optical, chow1985ring, bergh1984overview, rozelle2009hemispherical}.

However, commercial optical gyroscopes such as RLGs, HRGs and FOGs suffer from significant size and weight limitations that restrict their deployment in compact applications. While recent research has explored miniature navigation-grade gyroscopes using micro-resonators \cite{Liang:17, Miri2019, lai2019observation, Li:17, silver2021critical,Khial2018Nov}, these devices often require complex assembly procedures that hinder widespread adoption. The growing demand for miniature navigation-grade gyroscopes in drones, underwater vehicles, and satellites has created an urgent need for new approaches. Integrated photonic devices offer compelling advantages, including reduced device footprint, enhanced stability, and scalable manufacturing processes.

The fundamental challenge facing integrated photonic Sagnac gyroscopes stems from the direct relationship between the Sagnac effect and interferometric loop area \cite{Khial2018Nov, Li:17}. While fiber optic gyroscopes can incorporate hundreds or thousands of fiber loops to achieve extremely large effective areas (> 100 m$^2$ \cite{Nayak:11}),  integrated photonic devices are constrained to centimeter-scale dimensions on chip substrates. This inherent size limitation has prevented the widespread commercialization of integrated photonic gyroscopes. 


Researchers have investigated various approaches to amplify or enhance the Sagnac signal in chip-scale devices, including quantum optical and nonlinear optical techniques. For example, gyroscopes have been enhanced by quantum states of light, showing either reduced noise through squeezed states \cite{Xiao2020} or through superresolution with entangled photons \cite{Fink2019}. Counterpropagating Brillouin lasers have been used to measure rotation rates as low as 22~$^\circ/hr$ in a silica resonator of diameter 18~mm \cite{Li:17}. Moreover, creating an exceptional point by manipulating the gain and loss in two coupled modes can show orders of magnitude more sensitivity to perturbations such as rotation \cite{Miri2019,lai2019observation}. This phenomenon is closely related to level repulsion, or avoided crossing, in which two eigenvalues of a system diverge to avoid degeneracy \cite{Heiss2000}.

Recently, the Sagnac effect was also shown to be enhanced in a chip-scale microresonator by operating near the critical point of Kerr-induced spontaneous symmetry breaking \cite{silver2021critical}. 
Unfortunately, enhancement techniques that amplify the Sagnac effect simultaneously amplify various noise sources within the system, including laser phase and amplitude noise, as well as thermal fluctuations of the resonator \cite{zhang2019quantum,wang2020petermann,lau2018fundamental,Langbein2018,Khial2018Nov}. This concurrent amplification of thermal and amplitude noise represents a fundamental bottleneck limiting the performance of current chip-scale optical gyroscopes \cite{Silver:21}.

In this work, we demonstrate a chip-scale microresonator gyroscope that circumvents noise amplification by measuring the duration of the Kerr-induced symmetry-broken state as a laser scans in frequency across a resonance. We show that this approach achieves a $10^4$-fold amplification of the Sagnac signal while reducing average optical power in the resonator by 27 dB compared to continuous-wave pumping. This substantial power reduction proportionately decreases thermal noise contributions.

Another key innovation in our approach is suppressing noise from the slow drift of the resonance wavelengths over time from environmental fluctuations without the need for feedback control. We show that the start of the symmetry-broken state is dependent on the rotation rate, while the time at which the symmetry broken state ends is independent of the rotation. By using the duration of the symmetry broken state as the signal in our design, we eliminate the common mode noise from the resonator drift and see a 22 dB reduction in noise.

We demonstrate this technique using a 200~\textmu m silica microsphere, creating what we believe to be the smallest optical gyroscope reported to date. When normalized for resonator size, our gyroscope matches or exceeds the best reported values for resolution and angle random walk. This work addresses the fundamental noise-enhancement limitation that has prevented chip-scale optical gyroscopes from achieving practical performance levels using two key innovations: (1) Suppressing the thermal noise by 27 dB through lowering the average optical power and (2) Reducing noise from the slow resonance drift by 22 dB using the end of the symmetry broken state as a timing reference. By demonstrating that nonlinear enhancement can be achieved while simultaneously suppressing noise, we provide a pathway toward navigation-grade chip-scale gyroscopes. The technique is compatible with standard photonics fabrication processes and could enable integration with other inertial sensors for complete navigation systems on chip.

\section{Theoretical Background}


In an optical ring resonator of radius $R$ and refractive index $n$ rotating in its plane with an angular velocity $\Omega$ (in radians/sec), the Sagnac effect introduces a phase shift $\Delta \phi$ expressed as \cite{post1967sagnac}:
\begin{equation}
    \Delta \phi = \frac{8\pi}{\lambda c} \Omega A
    \label{eq:sagnac_quadratic}
\end{equation}
where $A$ is the area of the resonator and $\lambda$ is the vacuum wavelength of the pump. This causes the resonance frequencies of the counter-propagating modes of light to be offset by \cite{Malykin2014}
\begin{equation}
    \Delta \nu = \frac{2 R \Omega }{n \lambda},
    \label{eq:sagnac_shift}
\end{equation}

To contextualize this, with a 200~\textmu m resonator and a 1550 nm pump, the Sagnac frequency shift from Earth's rotation would be $7.2 mHz$. This shift is 7 orders of magnitude smaller than the resonance width of a state-of-the-art micro-resonator of Q factor $10^9$, which makes it extremely difficult to detect. The critical point associated with the Kerr-induced symmetry breaking in microresonators is one of several methods that have been proposed to enhance the Sagnac effect in chip-scale devices \cite{Kaplan:81, woodley2018universal,Silver:21,silver2021critical}.

The origin of the critical point in a bidirectionally pumped resonator with Kerr nonlinearity can be understood from a set of coupled equations (Eq. \ref{eq:Acw}) \cite{Kaplan:81, silver2021critical}. To illustrate the physical processes behind the symmetry breaking, suppose the resonator is pumped off-resonance in the blue-detuned side, with the counter-clockwise (CCW) mode having slightly more optical power due to an initial perturbation. Since the CCW mode has more power, the clockwise (CW) mode will experience greater cross-phase modulation, reducing the resonant frequency of the CW mode. This, in turn, further reduces the optical power in the CW mode. As the CW mode now has less power, the CCW mode will then experience less cross phase modulation, leading to an increase in its resonant frequency. This increase in CCW resonance frequency leads to increased optical power in the CCW mode, causing a positive feedback loop. With sufficiently high optical power in the resonator, the CW mode will be pushed completely out of resonance - we refer to this state as the \lq symmetry-broken state\rq. As a laser is tuned to the threshold of this symmetry breaking, the system shows a divergent behavior to external perturbations, such as a Sagnac shift. This divergent response when a resonator is continuously pumped close to the critical point was used by Silver et. al. \cite{Silver:21} to demonstrate a $10^4$ enhancement of the Sagnac shift in a 2.8 mm diameter silica microresonator.

The nonlinear gyroscope system can be described with the following system of differential equations \cite{zhu2019nonlinear, Zhu:20}:

\begin{align}
    \frac{d A_{\mathrm{cw}}}{dt} &= \left( -\left( \gamma_0+\gamma_{\mathrm{ex}} \right) - i \left( \Delta \omega_{\mathrm{cw}} - g + \Omega \right) \right) A_{\mathrm{cw}} - i g A_{\mathrm{ccw}} - i \frac{\eta}{\tau_0} B_{\mathrm{cw}}^{\mathrm{in}} \label{eq:Acw} \\
    \frac{d A_{\mathrm{ccw}}}{dt} &= \left( -\left( \gamma_0+\gamma_{\mathrm{ex}} \right) - i \left( \Delta \omega_{\mathrm{ccw}} - g - \Omega \right) \right) A_{\mathrm{ccw}} - i g A_{\mathrm{cw}} - i \frac{\eta}{\tau_0} B_{\mathrm{ccw}}^{\mathrm{in}} \label{eq:Accw} \\
    \frac{dT}{dt} &= -\frac{G}{H} \left(T-T_0\right)+\frac{\alpha c \tau_\mathrm{o}}{H} \left( \left| A_{\mathrm{cw}} \right|^2 + \left| A_{\mathrm{ccw}} \right|^2 \right) \label{eq:Temperature} \\
    \Delta \omega_{\mathrm{cw}}&=\omega_p-\omega_0\left(1-\frac{n_2}{n_0 A_{\mathrm{eff}}}\left(\left|A_{\mathrm{cw}}\right|^2+2\left|A_{\mathrm{ccw}}\right|^2\right)-\frac{dn}{dT} \frac{T}{n_0}\right)\label{eq:omega_cw} \\
    \Delta \omega_{\mathrm{ccw}}&=\omega_p-\omega_0\left(1-\frac{n_2}{n_0 A_{\mathrm{eff}}}\left(2\left|A_{\mathrm{cw}}\right|^2 + \left|A_{\mathrm{ccw}}\right|^2\right)-\frac{dn}{dT} \frac{T}{n_0}\right)\label{eq:omega_ccw}
\end{align}

\begin{equation}
\begin{aligned}
    P_{\mathrm{cw}}^{\mathrm{out}}  &= (B_{\mathrm{cw}}^{\mathrm{in}}-i \eta A_{\mathrm{cw}})^2 \\
    P_{\mathrm{ccw}}^{\mathrm{out}} &= (B_{\mathrm{ccw}}^{\mathrm{in}}-i \eta A_{\mathrm{ccw}})^2\end{aligned}
\label{eq:Pout}
\end{equation}

$A_{\mathrm{cw}}$ and $A_{\mathrm{ccw}}$ represent clockwise (CW) and counter-clockwise (CCW) coupled mode amplitude in the resonator; $B_{\mathrm{cw}}^{\mathrm{in}}$ and $B_{\mathrm{ccw}}^{\mathrm{in}}$ are the input light amplitudes; $\Omega$ is the angular frequency Sagnac shift from rotation; $T$ is temperature of the resonance mode volume; $T_0$ is room temperature; $\gamma_0$ is the intrinsic loss rate of resonator; $\gamma_{\mathrm{ex}}$ is the coupling loss rate; $g$ is the coupling rate between counter-propagating modes via backscattering; $n_2$ and $n_0$ are the nonlinear and linear indices of the resonator, respectively; $A_{\mathrm{eff}}$ is the effective cross-sectional mode area in the resonator; $\tau_0$ is the round-trip time of the resonator; $\eta = \sqrt{2\tau_0 \gamma_{\mathrm{ex}}}$ is the input coupling constant; $G$ and $H$ are the thermal conductance and thermal capacity of resonator; $\Delta \omega$ is the angular frequency detuning between a resonant mode (CW or CCW) and the pump frequency. In particular, equations \ref{eq:Acw} and \ref{eq:Accw} describe the growth of circulating power depending on the pump detuning and Sagnac shift, while equations \ref{eq:omega_cw} and \ref{eq:omega_ccw} determine the pump detuning from self and cross-phase modulation.


\begin{figure}[!h]
    \centering
    \includegraphics[width=0.8\linewidth]{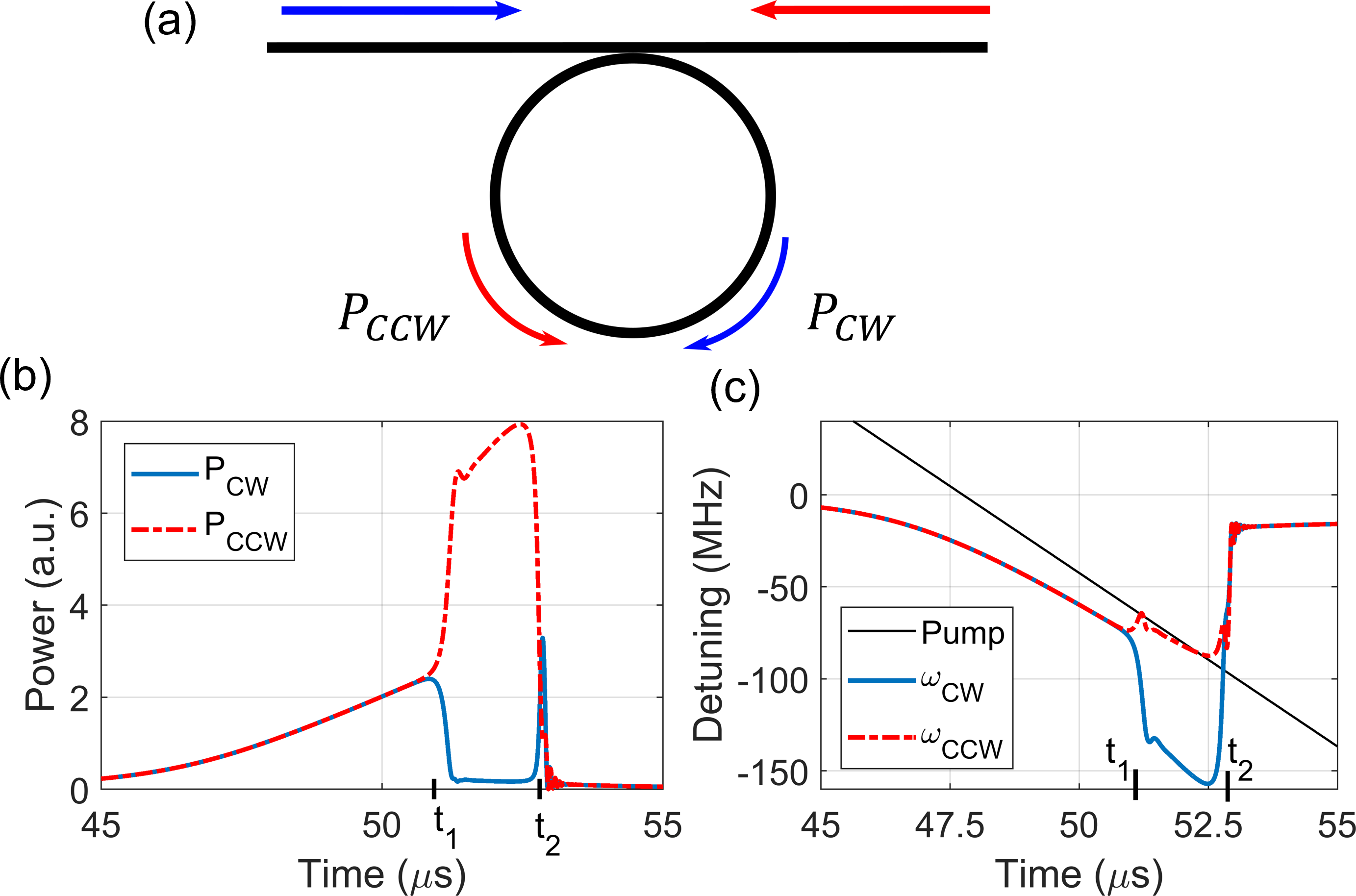}
    \caption{(a) Illustration of the Kerr symmetry breaking when scanning across a resonance. A continuously tunable laser pumps a microresonator equally in both directions. (b) Simulation showing the circulating power in both directions during a scan across a resonance. Circulating power is higher for the mode with resonant frequency closer to the pump frequency. At time $t_1$, the circulating power causes enough cross-phase modulation to shift the CW resonance away from the pump and the CCW resonance onto the pump, causing symmetry breaking. At time $t_2$ the pump frequency passes the CCW resonance and the effect is undone. (c) Simulation showing the same symmetry breaking process in terms of the frequency detuning of each resonance compared to the current pump frequency. The parameters used in this simulation are reported in the supplementary material.}
    \label{fig:fig1}
\end{figure}

In our \lq fast scanning\rq scheme, instead of tuning a laser to the threshold of the symmetry breaking, we scan the laser very quickly (at 0.33 MHz/\textmu s) across a resonance. Figure \ref{fig:fig1} illustrates the Kerr symmetry breaking during the laser scan in our gyroscope as it is rotating in the CCW direction. In this fast-scanning method the CW and CCW pumps start far blue-detuned from resonance and are simultaneously scanned across the resonance. As the pump light begins coupling into the resonator, it slightly favors the CCW mode due to the Sagnac effect. As depicted in Figure \ref{fig:fig1}(b), the system enters the symmetry broken state at time $t_1$. Eventually, the pump frequency matches that of the CCW resonance, and the resonator exits the symmetry-broken state at time $t_2$. As the pump leaves the CCW resonance, the CCW circulating power lowers, and the resonant frequencies of both CW and CCW modes shift back to their initial unperturbed positions.

These dynamics can also be understood by plotting the change in resonant frequencies over time. The Kerr-shifted resonant frequencies and the pump frequency are plotted in Figure \ref{fig:fig1}(c). As the optical power builds up in the resonator, the cross-phase modulation amplifies the original Sagnac frequency shift until the CW and CCW resonances are several MHz apart. 

The key innovation in our work is recognizing that $t_1$ is strongly dependent on the Sagnac shift, while $t_2$ is only very weakly dependent on the Sagnac shift. Furthermore, $t_1$ and $t_2$ have an equal dependence on the slow resonance drift due to environmental variations. Therefore, $(t_2-t_1)$ is strongly dependent on the Sagnac shift, and independent of the resonance drift from environmental variations. We experimentally measured a 22 dB reduction in noise when measuring rotation using $(t_2-t_1)$ vs. measuring $t_1$ against the clock signal driving the laser scanning.

The sensitivity of Kerr-enhanced optical gyroscopes is fundamentally constrained by multiple noise sources, with laser phase noise and thermal fluctuations representing the dominant limitations. While operation near exceptional points or critical points can enhance sensitivity, these regimes typically amplify noise proportionally with the signal enhancement \cite{Langbein2018, Silver:21}. The fast scanning methodology presented in this work mitigates both noise sources by minimizing the duration for which the laser remains tuned to resonant frequencies.

Compared to continuously locking to the side of a resonance of width $\Delta f$, our method of scanning across the resonance over a range $\Delta f_{\rm scan}$ reduces the average circulating power in the resonator by a factor of $\Delta f / \Delta f_{\rm scan}$.
This reduction corresponds to a 27 dB decrease in average power for our experimental configuration, yielding a proportional suppression of thermal noise. 


The minimum time to build up sufficient optical power for nonlinear effects in the resonator as the laser is scanned across a resonance places an upper limit on the laser scan rate. This upper limit was determined by solving the coupled differential equations describing the system. The simulations suggests that the upper limit on scan rate $\zeta$ scales empirically with the resonance linewidth $\Delta f$ as $\zeta < \frac{\Delta f}{3\tau}$, where $\tau = 2Q / \omega_0$ denotes the cavity time constant for exponential decay of circulating power. For our experimental parameters, this translates to an upper limit in the scan rate of $\zeta <$ 3 MHz/\textmu s. The implemented scan rate of 0.33 MHz/\textmu s was ultimately limited by the bandwidth of the proportional-integral-derivative (PID) control system employed for the tunable laser power stabilization.


\subsection{Backscattering}
Coupling between counter-propagating modes is a common problem for resonators, and it is most often caused by Rayleigh backscattering from material defects and geometric roughness in the resonator. 
The omnidirectional dipole radiation from these defects introduces coupling between the CW and CCW modes.
The effect of backscattering on optical gyroscopes has been studied for decades \cite{Iwatsuki:84, Etrich1992} and it is a common limiting factor in their performance \cite{Lloyd2013}.  The common description for this phenomenon is that two originally orthogonal modes, CW and CCW, are coupled into two hybrid modes with shifted resonance frequencies \cite{Iwatsuki:84}.  Thus, it becomes ambiguous whether any detected resonance shift is from the Sagnac effect or from the backscattering.

The Kerr-enhanced gyroscope faces a similar problem from backscattering, but from a different origin. Coupling between CW and CCW modes can reduce or entirely eliminate the nonlinear enhancement. This can be seen in equation \ref{eq:Acw} . Not only does the back-coupling term $g$ directly compete with the Sagnac shift $\Omega$, but the following term $-i g A_{\mathrm{ccw}}$ pulls both mode amplitudes $A_{\mathrm{ccw}}$ and $A_{\mathrm{cw}}$ toward a common value. This directly counteracts the positive feedback loop leading to the symmetry-broken state. Numerical simulations show that an increased backscattering rate increases the input power required to observe a bifurcation by a factor of  approximately $g/\gamma_0$. 



For our experiment, backscattering reduces the sensitivity of the gyroscope and increases the threshold power for symmetry breaking. The microspheres used in this experiment were fabricated by cleaving and melting a silica fiber under a high-voltage electric arc. The impurities are likely introduced by nanoparticles on the surface of the fiber that persisted after our ultrasonic cleaning process. Thus, we fabricated several such spheres, and tested each for high quality resonances ($Q > 10^8$) and low backscattering ( < -30 dB) by measuring both the transmitted and reflected power as the laser is scanned across a resonance \cite{PhysRevA.83.023803}. We then selected those that meet these threhsolds for the full gyroscope experiment. 

\section{Experiment}
\begin{figure}[ht]
    \centering
    \includegraphics[width=0.8\textwidth]{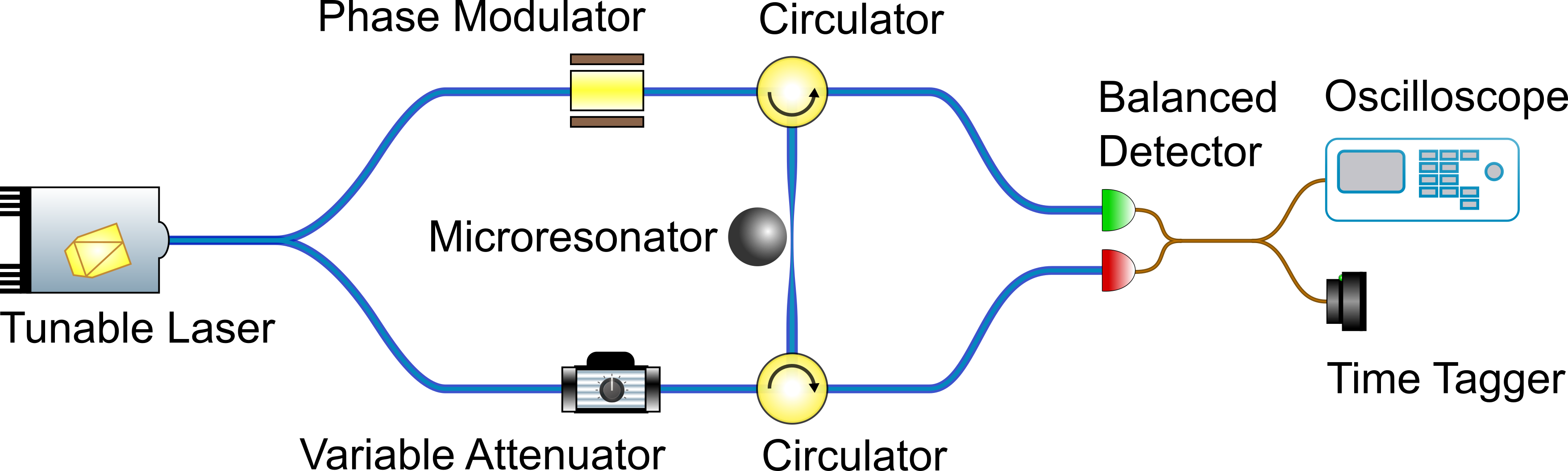}
    \caption{Schematic of a bidirectionally pumped microresonator gyroscope. A tapered fiber is used to couple light at 1550 nm into a 200 \textmu m Silica microsphere of Q factor $1.2\times10^8$. The laser is scanned across the resonance at 0.33 MHz/\textmu s  as the phase modulator is supplied with a voltage ramp to simulate a Sagnac shift. The Sagnac shift seeds a symmetry breaking between the CW and CCW modes of the resonator, which is enhanced by the Kerr nonlinearity. The balanced detector captures the symmetry breaking between the counterpropagating beams. The time tagger records the duration of the symmetry-broken state, which is a measure of the angular velocity. Two polarization controllers (not shown) allow us to ensure that the CW and CCW pumps are polarized along the same orientation.}
    \label{fig:schematic}
\end{figure}

The optical circuit used in the experiment is summarized in Fig. \ref{fig:schematic}. Light from a narrow-linewidth tunable diode laser (Toptica CTL 1550) is split into two arms of the circuit, which are coupled using a a silica tapered fiber (SMF-28) into the clockwise (CW) and counter-clockwise (CCW) modes of a 200 \textmu m diameter silica microsphere of quality factor $1.2 \times 10^8$. The silica fiber was tapered adiabatically to a width of about 900~nm by heating the bare fiber above the glass transition temperature and stretching from either side using motorized stages. An electronically controlled variable attenuator balances the input power into the microsphere from both directions. The difference in circulating powers is measured using a pair of balanced detectors. A time tagger records the duration of the bifurcation event, which is used as a measure of the rotation rate.

To reduce the mechanical complexity of the experiment, we simulate rotation (Sagnac shift) using a phase modulator (Thorlabs LN65S-FC) along one arm of the gyroscope. This is common practice in proof-of-concept optical gyroscope designs \cite{Silver:21}. A sawtooth signal to the phase modulator causes a small shift in the laser frequency in the corresponding arm, simulating the Sagnac effect. The timing of the sawtooth wave is carefully controlled so that the phase modulator does not apply an additional phase shift at the instant the resonator is optically probed. i.e, if the phase shift introduced by the modulator is $\phi$, then $\phi=0$ at the time of measurement, but $d\phi/dt = \delta f_\mathrm{Sagnac}$ can be independently tuned to simulate a range of Sagnac frequency shifts. This avoids the potential for any spurious effects arising from the phase modulator. The sawtooth waveform operates at 150~Hz and below 2~V, well below the bandwidth and $V_\pi$ of the phase modulator.

\subsection{Mitigating Interference Noise}
Due to the nature of fiber-coupled devices, small amounts of optical power is reflected from the optical components and fiber connections. This reflected power will interfere with the counter-propagating light, and alters the balance of CW and CCW pump powers seen by the resonator. This interference itself is not detrimental to the operation of the gyroscope, as the variable attenuator is able to balance the CW and CCW powers at the coupling point between the tapered fiber and the resonator.

However, any variation in the path length in either arm of the resonator will cause a change in the phase of the interference and the effective pump powers seen by the resonator. This variation affects the duration of the bifurcation event, leading to a loss in sensitivity of the gyroscope. To mitigate the effects of interference, we first minimize back-reflections by splicing the fiber directly to optical components wherever possible, and using angled (APC) connectors otherwise. We also introduce an extra 2.6 m of fiber in one arm of the circuit, causing the frequency of the CW beam to be offset from the CCW beam by about 4.3 kHz as we scan the laser at 0.33 MHz/\textmu s. We then compensate this frequency offset with a pump power difference so that there are equal amounts of circulating power in both CW and CCW modes in the resonator before bifurcation. 
This frequency offset reduces the temporal coherence between the two pumps, washing out any interference noise at long time scales \cite{Silver:21,woodley2018universal,silver2021critical,garbin2020asymmetric}.

These mitigations are likely unnecessary if the optical circuit is integrated into a monolithic chip-based platform. The short optical path lengths of such a device would ensure that there would be minimal variation in the phase of any interference. This passive stability would allow us to balance the circulating powers in the resonator using an inline variable attenuator at the time of calibration.


\section{Results and Discussion}

\begin{figure}[!h]
    \centering
    \includegraphics[width=0.8\textwidth]{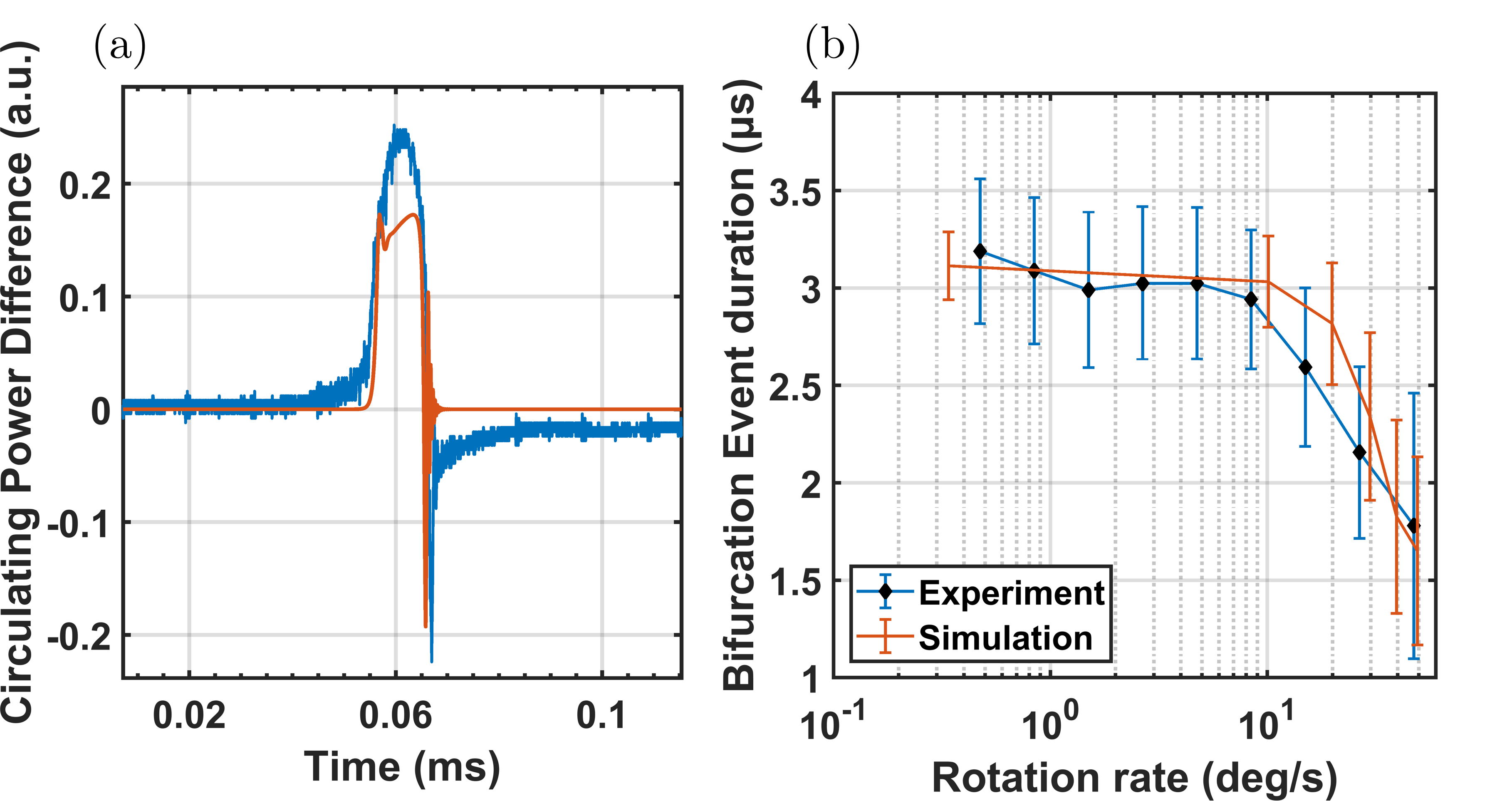}
    \vspace{+1em}
    \caption{(a) Time-domain signal showing the bifurcation event as the laser is scanned at 0.1 MHz/\textmu s, compared with the corresponding signal from a simulated bifurcation event. (b) Duration of the bifurcation event as a function of rotation rate as the laser is scanned at 0.33 MHz/\textmu s. Each data point is averaged across 150 samples. The data is compared with simulation data, taken by repeating the simulation in (a) with an added 12 kHz of frequency noise to the pump laser. Because the simulation introduces a random process, each data point represents the average and standard deviations from 40 repetitions of the simulation.}
    \label{fig:data}
\end{figure}

The duration of the symmetry broken state is quantified using a balanced detector that measures the difference in circulating powers in the CW and CCW modes. Figure \ref{fig:data}(a) shows the bifurcation event in the time domain. The rising and falling edges at a small positive threshold on the signal are respectively used to find the start and end of the bifurcation event using a time tagger. The duration of the bifurcation event has a clear dependence on the rotation rate, as seen in Figure \ref{fig:data}(b).

The sensitivity of microresonator gyroscopes have a quadratic dependence on the resonator radius (Equation \ref{eq:sagnac_quadratic}) \cite{Liang:17}. To compare the technique presented in this work to existing methods while controlling for the resonator sizes, we normalize the performance metrics — sensitivity and angle random walk (ARW) — by resonator area. This approach accounts for the inherent advantage of larger resonators and enables a direct comparison of different methods to measure the Sagnac effect in chip-scale devices. Achieving effective noise control allows our device to attain operational performance that matches or exceeds prior results, when adjusted for resonator size (Table 1).

\begin{table}[b]
    \begin{adjustwidth}{-2cm}{-2cm}
        \centering
        \setlength\extrarowheight{-2pt}
        \captionsetup{width=.85\linewidth}
        \caption{Comparing this work to previous reports of optical microresonator gyroscopes. As the Sagnac effect scales with the area of the resonator, normalizing the performance metrics by resonator area permits a direct comparison of different techniques to measure the Sagnac phase shift. The lower limit for the sensitivity reported in our work is the smallest rotation that has been detected, while the upper limit is the rotation rate that provides an easily detectable signal from a single measurement without averaging across multiple samples.}
        \label{table:comparison}
        \begin{tabularx}{1.1\textwidth} { |m{4cm} | >{\raggedright\arraybackslash}X | >{\raggedright\arraybackslash}X | >{\raggedright\arraybackslash}X | >{\raggedright\arraybackslash}X | >{\raggedright\arraybackslash}X | }
             \hline
             Authors & Sensitivity ($deg/s$, S) & ARW ($deg/\sqrt{Hr}$) & Resonator size (mm) &  Figure of merit: S*Area & Figure of merit: ARW*Area\\ 
             \hline
             Liang, Makeli, et al.\cite{Liang:17}       & 0.0008    & 0.02  & 7     & 0.032 & 0.76\\ 
             \textbf{This Work}                         & 0.9 - 6.0    & 28  & 0.20   & \textbf{0.028 - 0.188} & \textbf{0.88}\\
             Li, Vahala, et al.\cite{Li:17}             & 0.006     & 0.25  & 18    & 1.55  & 63.6\\
             Lai, Vahala, et al.\cite{lai2020earth}     & 0.001     & 0.07  & 36    & 1.01  & 69.1\\
             Khial, Hajimiri, et al.\cite{Khial2018Nov} & 6     & 650       & 1.41$^\dagger$  & 9.42   & 1021\\ 
             Silver, Del'Haye, et al.\cite{Silver:21}   & 2         & 400$^*$   & 2.8   & 12.3  & 2461\\ 
             \hline
        \end{tabularx}
    \end{adjustwidth}
    \small{
    $^*$ ARW has been estimated from noise values reported in the publication.\newline
    $^\dagger$ Khial, et al. used two 1 mm rings. For a reasonable performance normalization, we consider the gyroscope to have equivalent size to a $\sqrt{2}$~mm ring resonator.}
    \vspace{-5px}
\end{table}

In the absence of the Kerr enhancement, the Sagnac shift leads to an extremely small difference in circulating powers during the laser scan, as the laser first couples into the CCW mode and then to the CW mode. The duration of this pure Sagnac signal $s$ will be of the order of $s=\Delta \nu/\zeta$, where $\Delta \nu$ is the Sagnac shift from Eq. \ref{eq:sagnac_shift} and $\zeta$ is the laser scan rate. 
With a Kerr enhancement, we will see a symmetry breaking behavior and the duration of the symmetry broken state $(t_2-t_1)$ shows an enhanced response to the rotation. 
We experimentally observe that the duration of the symmetry-broken state $(t_2-t_1)$ is $5.2\times10^4$ times more sensitive to rotation than the pure Sagnac signal $s$. i.e, for a small change in the rotation rate $\partial \Omega$,
\begin{align*}
    \frac{\partial (t_2-t_1)}{\partial \Omega} &= 5.2 \times 10^4 \frac{\partial s}{\partial \Omega} \\
    \label{eq:nonlinear_enhancement}
\end{align*}

We include more details about the estimation of this enhancement factor in the supplementary material. This enhancement is comparable to the $10^4$ enhancement of the Sagnac signal reported by Silver et al. \cite{Silver:21} by tuning the laser to the threshold of the symmetry breaking and measuring the difference in optical power. However, due to the reduction in thermal and environmental noise from the fast scanning, we observe three orders of magnitude better area-normalized performance (Table \ref{table:comparison}). We note that this enhancement factor depends on the rotation rate, so we report this enhancement near the lowest rotation rate we are able to measure using our gyroscope.

The duration of the bifurcation event is shortest when the circulating powers are perfectly balanced. The bifurcation duration increases steeply with small deviations from the zero-bias point. In this experiment, there is a small residual power imbalance when the gyroscope is at rest because of the practical limits of the detectors and the variable attenuator. As we increase the rotation rate, the resonance in the direction with lower optical power at rest is pushed closer to the pump frequency, leading to less imbalance in the circulating power. Consequently, at higher rotation rates we observe a shorter bifurcation duration. 

Simulations indicate that this sensitivity to rotation is reasonable given the frequency noise in the pump. By including 12 kHz of frequency noise into the simulation, we obtained a distribution of bifurcation event durations shown in Figure \ref{fig:data}(b). Any additional discrepancy between experiment and theory is likely due to laser amplitude noise introduced during the scan.
Notably, in a noise-free scenario shown in the supplementary material, the sensor could be two orders of magnitude more sensitive to rotation. Thus, we speculate that a lower-noise laser and scanning method could vastly improve the gyroscope's sensitivity to rotation.

The performance of the gyroscope is characterized by the sensitivity, which is the smallest rotation that produces a measurable difference in output, and the angle random walk (ARW), which is a measure of how the reported angular orientation (pitch, yaw, roll) gradually drifts over time. The smallest rotation that is measured by our gyroscope is 0.9 $^\circ/s$. While we expect the bifurcation duration to be a monotonously decreasing function of rotation rate, we have observed that the variation in bifurcation duration is lower than the standard deviation until about 6 $^\circ/s$. We therefore report the sensitivity as between 0.9 - 6 $^\circ/s$. The ARW is measured using the technique described in the Supplementary material. The area-normalized sensitivity is 0.028-0.188 $^\circ mm^2/s$, which is competitive with the best reported value on a microresonator gyroscope \cite{Liang:17}. The area normalized ARW of 0.88 $^\circ mm^2/\sqrt{hr}$ is also comparable to the lowest reported value of 0.76 $^\circ mm^2/\sqrt{hr}$ \cite{Liang:17}.

This record-high performance in a sub-mm resonator was enabled by two concurrent innovations - A 27 dB reduction in average optical power from the fast-scanning approach, and a 22 dB reduction in noise from environmental fluctuations by using the end of the symmetry-broken state as a timing reference.

\section{Conclusion}

We have demonstrated a novel way to enhance the Sagnac effect in an optical microresonator. The proof-of-concept demonstration in this work achieves a $5.2 \times 10^4$ enhancement of the Sagnac phase shift, via nonlinear enhancement using the Kerr effect. Notably, we enhance the Sagnac effect without proportionally increasing the noise floor. The fast-scanning mechanism allows us to suppress thermal noise by 27 dB, and the environmental noise by 22 dB. Our proof-of-concept demonstration achieves exceptional sensitivity and low noise in a small package.

There are several avenues for improvements as part of future efforts. The sensitivity of the gyroscope is limited by the imbalance of power between the clockwise and counter-clockwise pumps. Implementing feedback control for power stabilization is likely to improve the performance of the gyroscope significantly. Implementing the design on a photonic integrated platform will lead to additional gains in long-term stability and scalability. In addition, by increasing the diameter of the resonator, rotation sensitivities approaching those of commercial RLGs or FOGs on a small footprint are within reach using this proposed technique.

\begin{backmatter}
\bmsection{Funding}
This work was supported by the National Science Foundation (Grant number ECCS 2224065)

\bmsection{Acknowledgment}
The authors gratefully acknowledge James Erikson, Bright Lu and Mo Zohrabi for technical discussions. 

\bmsection{Disclosures}
The authors declare no conflicts of interest.

\bmsection{Data availability} Data underlying the results presented in this paper are available upon reasonable request.

\bmsection{Supplemental document}
The Supplemental document contains information about the angle random walk measurement, sensitivity enhancement calculations and simulation parameters.

\end{backmatter}


\bibliography{references}

\begin{thebibliography}{1}
\newcommand{\enquote}[1]{``#1''}

\bibitem{miller2012probability}
S.~Miller, \emph{Probability and random processes: With applications to signal processing and communications} (Academic Press, 2012).

\end{thebibliography}


\begin{thebibliography}{10}
\newcommand{\enquote}[1]{``#1''}

\bibitem{post1967sagnac}
E.~J. Post, \enquote{Sagnac effect,} {\protect\JournalTitle{Reviews of Modern Physics}} \textbf{39}, 475--493 (1967).

\bibitem{yang2025micro}
Z.~Yang, Y.~Deng, J.~Su, \emph{et~al.}, \enquote{From micro‐optical to quantum‐enhanced gyroscopes: A comprehensive review,} {\protect\JournalTitle{Laser \& Photonics Reviews}} \textbf{19}, 2402065 (2025).

\bibitem{arianfard2023sagnac}
H.~Arianfard, S.~Juodkazis, D.~J. Moss, and J.~Wu, \enquote{Sagnac interference in integrated photonics,} {\protect\JournalTitle{Applied Physics Reviews}} \textbf{10} (2023).

\bibitem{chow1985ring}
W.~Chow, J.~Gea-Banacloche, L.~Pedrotti, \emph{et~al.}, \enquote{The ring laser gyro,} {\protect\JournalTitle{Reviews of Modern Physics}} \textbf{57}, 61--104 (1985).

\bibitem{bergh1984overview}
R.~Bergh, H.~Lefevre, and H.~Shaw, \enquote{An overview of fiber-optic gyroscopes,} {\protect\JournalTitle{Journal of Lightwave Technology}} \textbf{2}, 91--107 (1984).

\bibitem{rozelle2009hemispherical}
D.~M. Rozelle, \enquote{The hemispherical resonator gyro: From wineglass to the planets,} in \emph{Proc. 19th AAS/AIAA Space Flight Mechanics Meeting,}  vol. 134 (2009), pp. 1157--1178.

\bibitem{oho2002optical}
S.~Oho, H.~Kajioka, and T.~Sasayama, \enquote{Optical fiber gyroscope for automotive navigation,} {\protect\JournalTitle{IEEE Transactions on vehicular technology}} \textbf{44}, 698--705 (2002).

\bibitem{Liang:17}
W.~Liang, V.~S. Ilchenko, A.~A. Savchenkov, \emph{et~al.}, \enquote{Resonant microphotonic gyroscope,} {\protect\JournalTitle{Optica}} \textbf{4}, 114--117 (2017).

\bibitem{Miri2019}
M.-A. Miri and A.~Al{\`{u}}, \enquote{Exceptional points in optics and photonics,} {\protect\JournalTitle{Science}} \textbf{363} (2019).

\bibitem{lai2019observation}
Y.-H. Lai, Y.-K. Lu, M.-G. Suh, \emph{et~al.}, \enquote{Observation of the exceptional-point-enhanced sagnac effect,} {\protect\JournalTitle{Nature}} \textbf{576}, 65--69 (2019).

\bibitem{Li:17}
J.~Li, M.-G. Suh, and K.~Vahala, \enquote{Microresonator brillouin gyroscope,} {\protect\JournalTitle{Optica}} \textbf{4}, 346--348 (2017).

\bibitem{silver2021critical}
J.~M. Silver, K.~T. Grattan, and P.~Del'Haye, \enquote{Critical dynamics of an asymmetrically bidirectionally pumped optical microresonator,} {\protect\JournalTitle{Physical Review A}} \textbf{104}, 043511 (2021).

\bibitem{Khial2018Nov}
P.~P. Khial, A.~D. White, and A.~Hajimiri, \enquote{Nanophotonic optical gyroscope with reciprocal sensitivity enhancement,} {\protect\JournalTitle{Nature Photonics}} \textbf{12}, 671--675 (2018).

\bibitem{Nayak:11}
J.~Nayak, \enquote{Fiber-optic gyroscopes: from design to production,} {\protect\JournalTitle{Applied Optics}} \textbf{50}, E152--E161 (2011).

\bibitem{Xiao2020}
X.-Q. Xiao, E.~S. Matekole, J.~Zhao, \emph{et~al.}, \enquote{Enhanced phase estimation with coherently boosted two-mode squeezed beams and its application to optical gyroscopes,} {\protect\JournalTitle{Physical Review A}} \textbf{102}, 022614 (2020).

\bibitem{Fink2019}
M.~Fink, F.~Steinlechner, J.~Handsteiner, \emph{et~al.}, \enquote{Entanglement-enhanced optical gyroscope,} {\protect\JournalTitle{New Journal of Physics}} \textbf{21}, 053010 (2019).

\bibitem{Heiss2000}
W.~D. Heiss, \enquote{Repulsion of resonance states and exceptional points,} {\protect\JournalTitle{Phys. Rev. E}} \textbf{61}, 929--932 (2000).

\bibitem{zhang2019quantum}
M.~Zhang, W.~Sweeney, C.~W. Hsu, \emph{et~al.}, \enquote{Quantum noise theory of exceptional point amplifying sensors,} {\protect\JournalTitle{Physical review letters}} \textbf{123}, 180501 (2019).

\bibitem{wang2020petermann}
H.~Wang, Y.-H. Lai, Z.~Yuan, \emph{et~al.}, \enquote{Petermann-factor sensitivity limit near an exceptional point in a brillouin ring laser gyroscope,} {\protect\JournalTitle{Nature communications}} \textbf{11}, 1610 (2020).

\bibitem{lau2018fundamental}
H.-K. Lau and A.~A. Clerk, \enquote{Fundamental limits and non-reciprocal approaches in non-hermitian quantum sensing,} {\protect\JournalTitle{Nature communications}} \textbf{9}, 4320 (2018).

\bibitem{Langbein2018}
W.~Langbein, \enquote{No exceptional precision of exceptional-point sensors,} {\protect\JournalTitle{Physical Review A}} \textbf{98}, 023805 (2018).

\bibitem{Silver:21}
J.~M. Silver, L.~D. Bino, M.~T.~M. Woodley, \emph{et~al.}, \enquote{Nonlinear enhanced microresonator gyroscope,} {\protect\JournalTitle{Optica}} \textbf{8}, 1219--1226 (2021).

\bibitem{Malykin2014}
G.~B. Malykin, \enquote{Sagnac effect in ring lasers and ring resonators. how does the refractive index of the optical medium influence the sensitivity to rotation?} {\protect\JournalTitle{Physics-Uspekhi}} \textbf{57}, 714--720 (2014).

\bibitem{Kaplan:81}
A.~E. Kaplan and P.~Meystre, \enquote{Enhancement of the sagnac effect due to nonlinearly induced nonreciprocity,} {\protect\JournalTitle{Optics Letters}} \textbf{6}, 590--592 (1981).

\bibitem{woodley2018universal}
M.~T. Woodley, J.~M. Silver, L.~Hill, \emph{et~al.}, \enquote{Universal symmetry-breaking dynamics for the kerr interaction of counterpropagating light in dielectric ring resonators,} {\protect\JournalTitle{Physical Review A}} \textbf{98}, 053863 (2018).

\bibitem{zhu2019nonlinear}
J.~Zhu, M.~Zohrabi, K.~Bae, \emph{et~al.}, \enquote{Nonlinear characterization of silica and chalcogenide microresonators,} {\protect\JournalTitle{Optica}} \textbf{6}, 716--722 (2019).

\bibitem{Zhu:20}
J.~Zhu, M.~Zohrabi, K.~Bae, \emph{et~al.}, \enquote{Nonlinear characterization of silica and chalcogenide microresonators: erratum,} {\protect\JournalTitle{Optica}} \textbf{7}, 185--185 (2020).

\bibitem{Iwatsuki:84}
K.~Iwatsuki, K.~Hotate, and M.~Higashiguchi, \enquote{Effect of rayleigh backscattering in an optical passive ring-resonator gyro,} {\protect\JournalTitle{Applied Optics}} \textbf{23}, 3916--3924 (1984).

\bibitem{Etrich1992}
C.~Etrich, P.~Mandel, R.~Centeno~Neelen, \emph{et~al.}, \enquote{Dynamics of a ring-laser gyroscope with backscattering,} {\protect\JournalTitle{Physical Review A}} \textbf{46}, 525--536 (1992).

\bibitem{Lloyd2013}
S.~W. Lloyd, M.~J.~F. Digonnet, and {Shanhui Fan}, \enquote{Modeling coherent backscattering errors in fiber optic gyroscopes for sources of arbitrary line width,} {\protect\JournalTitle{Journal of Lightwave Technology}} \textbf{31}, 2070--2078 (2013).

\bibitem{PhysRevA.83.023803}
X.~Yi, Y.-F. Xiao, Y.-C. Liu, \emph{et~al.}, \enquote{Multiple-rayleigh-scatterer-induced mode splitting in a high-q whispering-gallery-mode microresonator,} {\protect\JournalTitle{Physical Review A}} \textbf{83}, 023803 (2011).

\bibitem{garbin2020asymmetric}
B.~Garbin, J.~Fatome, G.-L. Oppo, \emph{et~al.}, \enquote{Asymmetric balance in symmetry breaking,} {\protect\JournalTitle{Physical Review Research}} \textbf{2}, 023244 (2020).

\bibitem{lai2020earth}
Y.-H. Lai, M.-G. Suh, Y.-K. Lu, \emph{et~al.}, \enquote{Earth rotation measured by a chip-scale ring laser gyroscope,} {\protect\JournalTitle{Nature Photonics}} \textbf{14}, 345--349 (2020).

\end{thebibliography}

\end{document}


\maketitle

\section{Calculation of the Angle Random Walk (ARW)}

\begin{figure}[htb]
    \centering
    \includegraphics[width=0.8\textwidth]{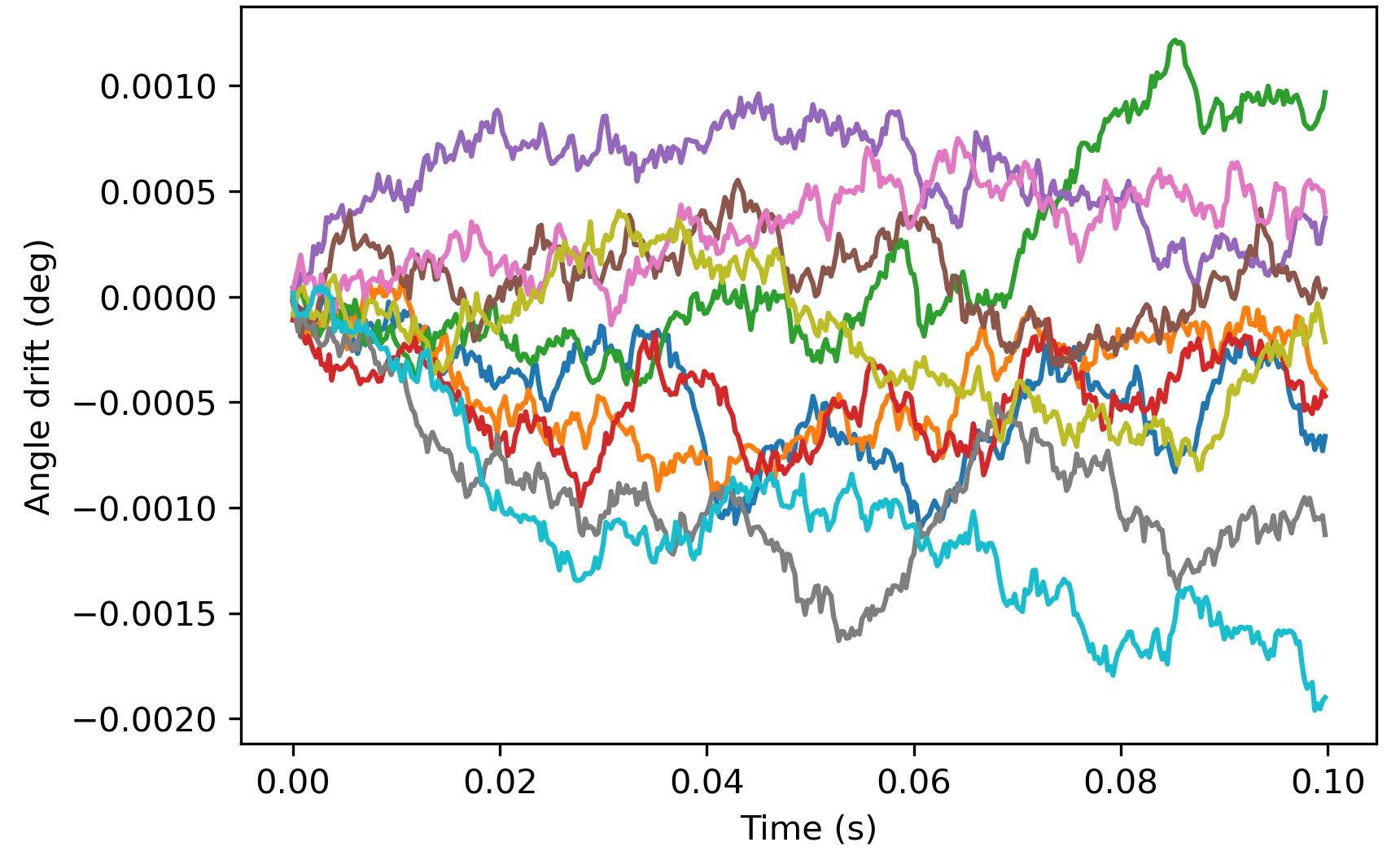}
    \caption{Simulated angle drift from white noise in the raw gyroscope reading for 10 separate datasets. The gyroscope is sampled at a fixed 4000 Hz. The noise has a flat power spectral density and is sampled from a Gaussian probability distribution with standard deviation 0.185 $\deg/s$. The angle drift is calculated by integrating the noise over time.}
    \label{fig:angle_drift}
\end{figure}

Gyroscope noise is typically characterized through their angle random walk (ARW). It is measured in units of $angle/\sqrt{time}$ - for example, $\degree/\sqrt{s}$ or $\degree/\sqrt{hr}$. This parameter is especially important for inertial navigation applications, where the output of the gyroscope is integrated over time to calculate the total angle of rotation. As the name suggests, ARW characterizes how much 'random walk' we can expect when the total angle of rotation is calculated by integrating the gyroscope reading over some given time.

Suppose the sensor is stationary. The true angular velocity is $0 \deg/s$, but we will measure non-zero readings from the sensor noise. Now suppose the noise in the system is predominantly white noise, i.e., the noise has a flat spectrum in the Fourier space. The integral of this white noise is called a Wiener process \cite{miller2012probability}. A Wiener process has two interesting properties:

\begin{enumerate}
    \item $E[W_t]=0$, i.e., the expectation value of a Wiener process is zero.
    \item $\sigma^2[W_t] \propto t$, i.e., the variance of the Wiener process is proportional to the duration of the Wiener process.
\end{enumerate}

For a gyroscope, the above properties have the following implications. Suppose we integrate the reading from the gyroscope over a fixed time $t$ to obtain the drift angle $A_t$. Further, assume that we do this measurement of drift angle multiple times to form a statistics for $A_t$.

\begin{enumerate}
    \item The average value of all measured $A_t$ values will be close to zero.
    \item The variance of these angle measurements will be proportional to the integration time. Or in other words, the standard deviation of the observed $A_t$ values will be proportional to $\sqrt{t}$
\end{enumerate}

The constant of proportionality between the standard deviation of the $A_t$ values and $\sqrt{t}$ is the angle random walk. While we may take a handful of datasets for some fixed integration time $t$ and find the AWR by dividing the variance of observer $A_t$ values by $\sqrt{t}$, there is a way to utilize the same amount of information to arrive at an estimate for ARW that better predicts the system dynamics. We describe the method below:

\begin{figure}[tb]
    \centering
    \includegraphics[width=0.8\textwidth]{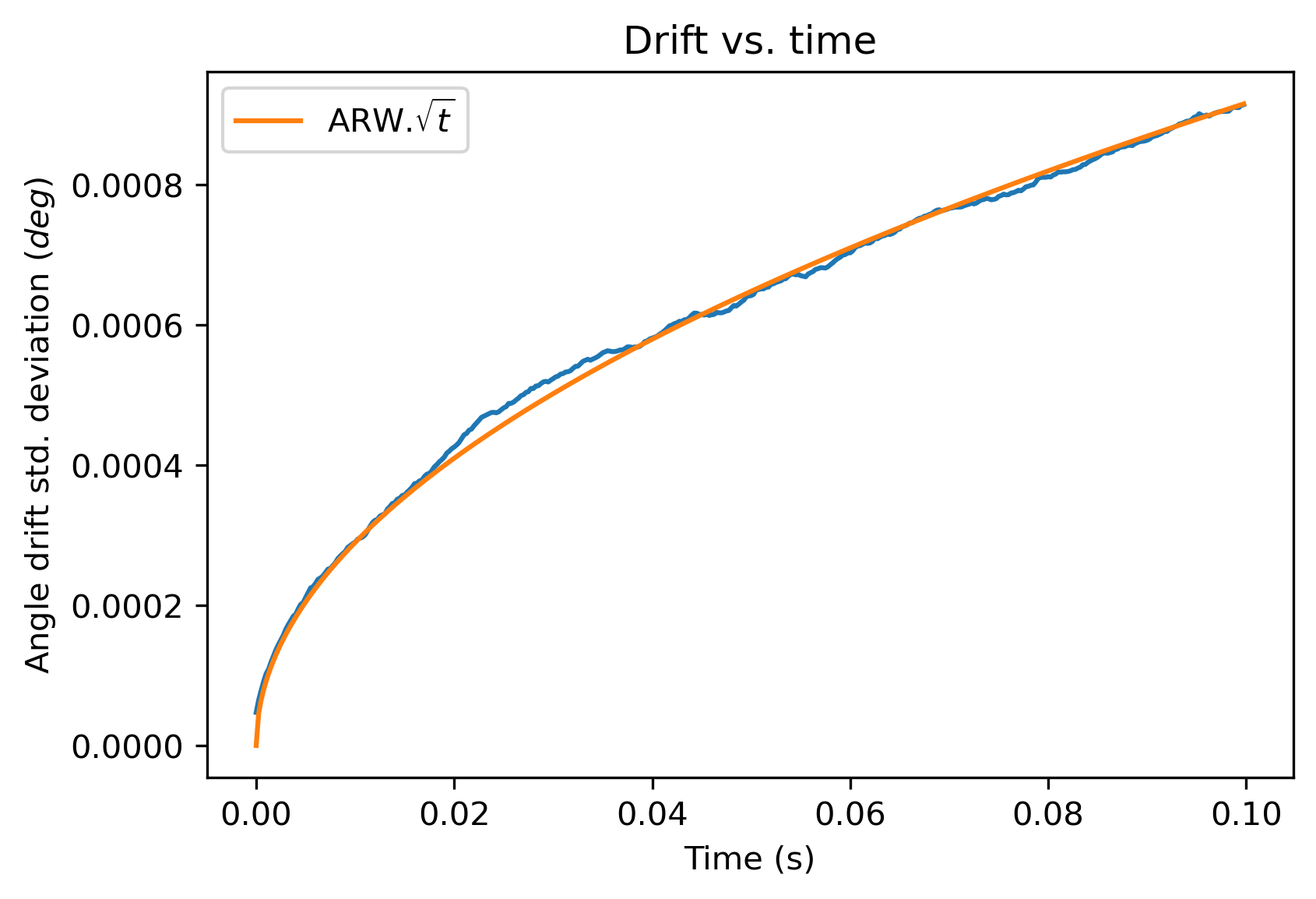}
    \caption{The standard deviation of the angle drifts over time is proportional to $\sqrt{t}$. The standard deviation in this example is calculated across 1000 datasets, each containing 400 samples. The constant of proportionality is the angle random walk.}
    \label{fig:ARW_measurement}
\end{figure}

\begin{enumerate}
    \item Collect readings from the gyroscope at rest, at equal intervals for as long as the system permits before the bias points of the system suffer from long-term drift. Suppose we collected N discrete samples.
    \item Divide up the time-series information into $m$ segments of equal duration, each containing $n=floor(N/m)$ samples. It is ideal to factorize N into $m\times n$, with both $m$ and $n$ as close to $\sqrt{N}$ as possible to minimize discarding samples.
    \item Obtain the angle drift over time for each of these sub-segments. The angle drift is calculated as the cumulative sum of the samples in the sub-segments multiplied by the sampling interval (Fig \ref{fig:angle_drift}).
    \item Fit a curve of the form to $y=B\sqrt{x}$ to the angle drift standard deviation vs. time curve (Fig \ref{fig:ARW_measurement}). This constant of proportionality $B$ is the angle random walk. 
\end{enumerate}

\section{Modeling Resonator Dynamics}


We numerically integrated these equations with a Runge-Kutta method with the input parameters $Q_0 = Q_{ex} = 1.2 \times 10^8$, $R = 100 \times 10^{-6} m$, $n_2 = 2.4 \times 10^{-20} m^2/W$, $n_0 = 1.3$, $\alpha = 1 \times 10^{-4} m^{-1}$, $g = 1 \times 10^5 Hz$, $A_{eff} = 2.5 \times 10^{-12} m^2$, $dn/dT = 9 \times 10^{-6} K^{-1}$, $H = \rho 2\pi R A_{eff} H_s$ where $H_s = 700 J / kg / K$ and $\rho = 2400 kg / m^3$, $\lambda_0 = 1.55 \times 10^-6 m$, $P_{in} = 125 \times 10^{-6} W$. Additionally, the scan started at $\omega_i -\omega_0 = 9 \times 10^8 $ Hz, $d\omega_{p}/dt = 6\pi \times 10^{12} Hz/s$ , and ran with $t_{\rm step} = $ 1 ps  over a duration of $t_{\rm total} = $  100 $\mu s$.  

\begin{figure}[h]
    \centering
    \includegraphics[width=0.5\linewidth]{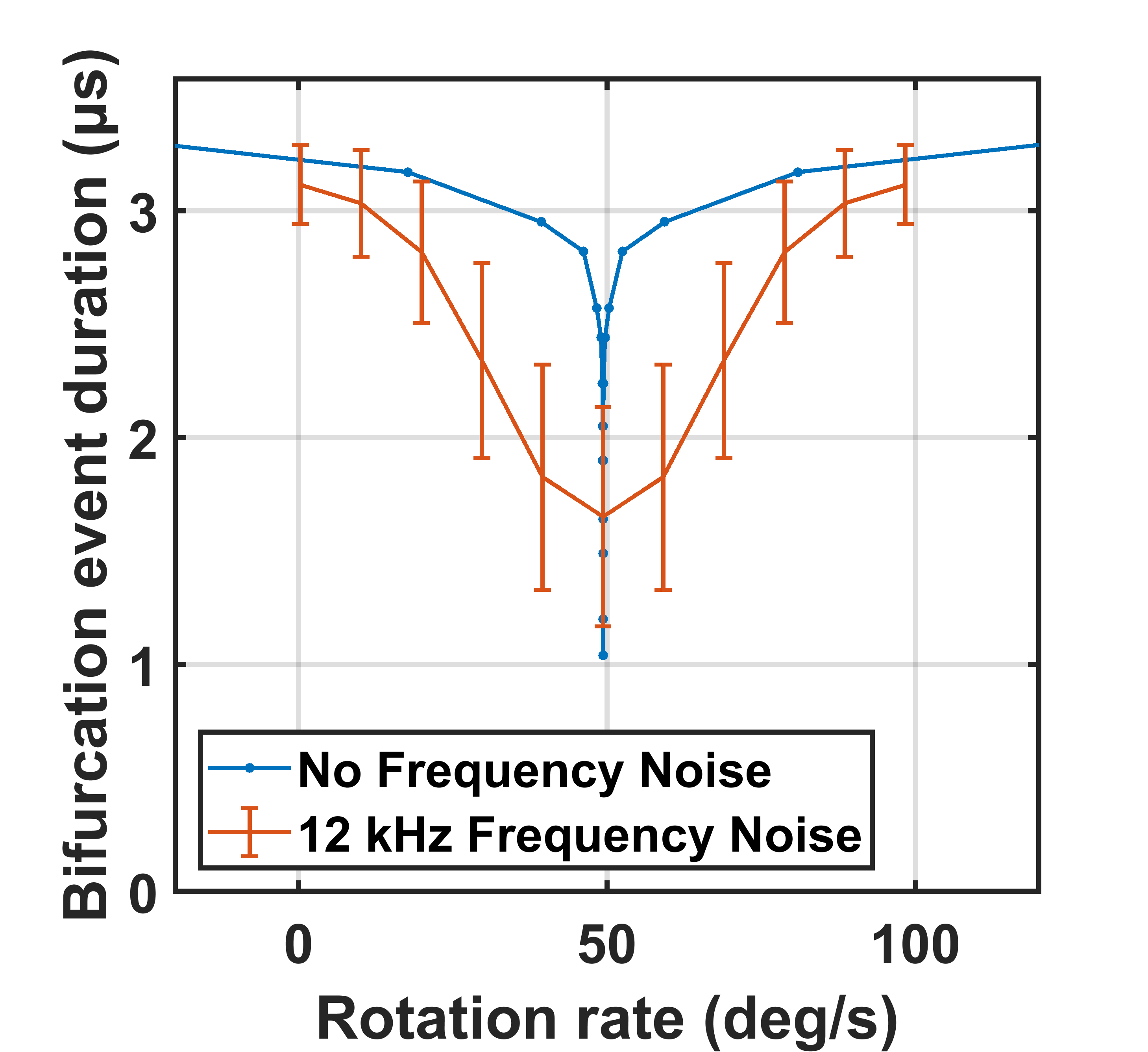}
    \caption{Comparison of the simulated bifurcation event duration with respect to rotation rate, with and without frequency noise. The relative width of each signal is 0.7 deg/s without noise, and 40 deg/s including noise.}
    \label{fig:simulated-rotation-sensitivity}
\end{figure}

Additionally, $12$ kHz of laser frequency noise was included by adding a random phase onto a Gaussian envelope function, taking the real part of its Fourier transform, and adding that to the pump frequency. A comparison of simulations with and without frequency noise is shown in Figure \ref{fig:simulated-rotation-sensitivity}. The added frequency noise effectively broadens the distribution by nearly two orders of magnitude, better matching the experimental data from Figure 3(b). The 12 kHz added noise is comparable to the 10 kHz linewidth of the Toptica CTL 1550 laser we used for this experiment.

\section{Calculating the Sensitivity enhancement}

From 0.47 deg/s to 0.84 deg/s, we observe a change in $(t_2-t_1)$ from 3.189~\textmu s to 3.088~\textmu s. Therefore,

\begin{align*}
    \frac{\partial (t_2-t_1)}{\partial \Omega} &= \frac{(3.189-3.088)\times 10^{-6}}{(0.84-0.47)\times \frac{\pi}{180}} \\ \\
    &= 1.56 \times 10^{-5} \\ \\
    \frac{\partial s}{\partial \Omega} &= \frac{\partial}{\partial \Omega} \left( \frac{\Delta \nu}{\zeta} \right) \\ \\ 
    &= \frac{\partial}{\partial \Omega} \left( \frac{2R\Omega}{n\lambda} \frac{1}{\zeta} \right)= \frac{2R}{n\lambda \zeta} \\ \\ 
    &= \frac{2\cdot 100 \times 10^{-6}}{1.3 \cdot 1.55\times 10^{-6} \cdot 0.33\times 10^{12}} \\ \\ 
    &= 3.00\times 10^{-10}
\end{align*}

The enhancement factor is therefore $1.56 \times 10^{-5}/ 3.00\times 10^{-10}=5.2\times10^4$. Note that $\partial s/\partial \Omega$ is calculated analytically and depends only on the resonator radius, the wavelength, effective index and scan rate.

\bibliography{references}